\documentclass[superscriptaddress,amssymb,amsmath,nofootinbib,twocolumn,aps,prd,longbibliography]{revtex4-1}
\usepackage[utf8]{inputenc}
\usepackage{mathtools}
\usepackage{hyperref}
\usepackage{slashed}
\hypersetup{colorlinks=true, urlcolor=blue,linkcolor=blue, citecolor=blue}
\usepackage{longtable}
\usepackage{graphicx}

\begin{document}
\title{On the permanence of renormalons in compactified spaces}
\author{E. Cavalcanti}
\email[]{erich@cbpf.br}
\affiliation{Centro Brasileiro de Pesquisas F\'{\i}sicas/MCTI, 22290-180 Rio de Janeiro, RJ, Brazil}

\begin{abstract}
We investigate the existence and behavior of renormalon singularities with respect to $d$ spatial compactifications and quasiperiodic boundary conditions. Employing a toy model (scalar field theory with quartic interaction) we find that the size of the compactification and the number of compactified dimensions do not influence the number and the location of the renormalon poles. The only influence occurs in the residues. We enforce the need to carefully check the asymptotic approximations every time there is some result about the appearance of new renormalon poles or their cancellation.
\end{abstract}

\maketitle
 

\section{Introduction}
\label{sec:intro}

The employment of perturbative techniques under some parameter ($g$, for example) to consider a difficult problem is as old as physics itself. However, one often runs into the obstacle that the perturbative series on $g$ lacks information about the nonperturbative phenomena of the theory under interest. It happens because the value of the parameter $g$ might fall beyond convergence radius or -- more dramatically, but usual -- because the perturbative series has no convergence radius at all and is asymptotic. 

Usually, many perturbative series turns out to have a factorial growth in their coefficients, making them explicitly divergent. Therefore, we must understand the asymptotic series as a formal series that acts as a representation of the \emph{exact} solution and that one can extract from it the nonperturbative behavior. From the formulation of asymptotic series by Poincaré, we know that, although the perturbative series diverge, we can use optimal truncation (stop the summation at some optimal term) to approach the exact solution. Also, the remaining of the sum -- the error in the truncation -- is sometimes also calculable. For phenomenological purposes in quantum field theory, condensed matter, and even quantum mechanics, it is common to sum the perturbative series up to its optimal truncation. Advanced techniques are necessary to go beyond this.

A standard methodology to overcome the actual divergence of the perturbative series is to use a Borel summation. That is, the original series -- in the coupling-constant plane --  is related to a new series -- that lives in, what we call, the Borel plane -- that controls the factorial growth and is, perhaps, summable. Then, we employ a Borel inverse transform to return to the original coupling-constant plane. If the inverse Borel transform for the problem is well defined, we achieve a nonperturbative solution out of the perturbative series.

However, it turns out that some sources of divergence in the perturbative series appear as singularities in the Borel plane and can also prevent us from employing the inverse Borel transform. In quantum field theories, some familiar sources of divergence are the instanton and renormalon singularities. The instanton problem has its root in the factorial growth in the number of Feynman diagrams at each new order in perturbation theory. The renormalon problem is more subtle and arises due to an infrared (IR) or ultraviolet (UV) divergence in the integration of the internal momentum of a set of diagrams. 

The renormalon phenomenon is an indication of the range of validity of the original perturbative series. It became a convenient source even to phenomenological applications, as it indicates the need for higher corrections (For some applications in the last years check refs.~\cite{Cvetic:2019jmu,Hadjimichef:2019vbb,Takaura:2019vak,Hayashi:2019mlb,Corcella:2019tgt,Maiezza:2019dht,Nason:2019gff,Cvetic:2018qxs,Ortega:2018oae,Kataev:2018mob,Mateu:2018zym,Boito:2018dmm,FerrarioRavasio:2018ubr,Brambilla:2018tyu,Braun:2018brg,Kataev:2018fvx,Bell:2018gce,Takaura:2018lpw,Takaura:2018vcy,Suzuki:2018vfs,DelDebbio:2018ftu,Kataev:2018gle,Peset:2018ria,Bazavov:2018omf,Brambilla:2017hcq,Nason:2017cxd,Ortega:2017lyb,Mateu:2017hlz,Hoang:2017btd,Ahmadov:2017fcf,Steinhauser:2016xkf,Jamin:2016ihy,Ayala:2016sdn,Mishima:2016vna}). In a modern sense, the renormalon problem indicates that the original asymptotic perturbative series needs a \emph{transseries} contribution~\cite{Anber:2014sda,Dunne:2016nmc,Maiezza:2018ags,Marino:2019wra,Marino:2019eym,Marino:2019fvu,Ishikawa:2019oga,Dondi:2020qfj} (series that contains \emph{transmonomials}, as $x^n e^{mx} \ln^o x$, instead of usual monomials, as $x^n$ ). The resurgence program introduces powerful and novel techniques beyond the classic approach to asymptotic series, managing -- in principle -- to relate perturbative and nonperturbative sectors of a theory. Although it lacks the discussion of the recent years on the resurgence program, ref.~\cite{Beneke:1998ui} is still a comprehensive review on the topic of renormalons. See also refs.~\cite{Brambilla:1999xf,Brambilla:2004jw,Melnikov:2000qh,Bigi:1994em,Beneke:1994sw,Mueller:1984vh,Sumino:2000tk,Brambilla:2000cs} for more details.

In recent years, there has been a growing excitement in the field due to investigations of nonperturbative aspects of field theory based on its perturbative expansion, both to clarify and to apply the new techniques and assumptions. One crucial inquiry is whether we genuinely recover the correct nonperturbative solution out of the perturbative series. A simple proposal to elucidate this point is to compare it with a theory with a known exact solution. However, there are just a few known solvable models in field theory with renormalon singularities~\cite{Marino:2020dgc}. 
For other theories, it is useful to consider some methodology to cure the renormalon singularities at some regime. In this sense, some authors propose that we can cure the renormalon ambiguities if the original space is compactified ($\mathbb{R}^D \rightarrow\mathbb{R}^{D-1}\times \mathbb{S}_L^1$). The so-called adiabatic continuity conjecture (or just continuity conjecture)~\cite{Kovtun:2007py,Shifman:2008ja, Unsal:2008ch} states that if the renormalon vanishes in the compactified scenario and we can move the length parameter continuously (from small $L$ to large $L$) without undergoing a phase transition, then we can relate the renormalon-free compactified theory (at small-$L$) with the original theory (at very large $L$, the bulk). 

One popular proposal is that in a compactified space some semi-classical structures as fractional instantons, bions (the composition of fractional instantons), or molecules of bions arise and might cancel out the renormalon ambiguities driving the theory well-defined~\cite{Kovtun:2007py,Dunne:2012zk,Argyres:2012vv,Dunne:2012ae,Argyres:2012ka}. However, there are some indications that these semi-classical solutions do not correspond to the renormalon ambiguities~\cite{Cherman:2013yfa,Ishikawa:2019tnw,Morikawa:2020agf,Dunne:2020gtk}. Recently, ref.~\cite{Cherman:2013yfa,Ishikawa:2019tnw,Morikawa:2020agf,Dunne:2020gtk,Unsal:2020yeh} reinvestigated the idea of cancellation of the renormalon ambiguities. 

Another perspective that might take place is the algebraic cancellation of renormalon in the direct limit of dimensional reduction ($L\rightarrow0$)~\cite{Anber:2014sda,Misumi:2014raa,Nitta:2015tua,Cavalcanti:2018thz,Ashie:2020bvw}; it presupposes that we can smoothly remove one dimension ($D \rightarrow D-1$) and that with one less spatial direction the renormalon problem does not appear. With just one spatial restriction this is equivalent (except to the freedom regarding the boundary condition) to thermal quantum field theory, which is well-known and extensively studied due to the interest, for example, in thermal quantum chromodynamics (QCD). In the thermal QFT scenario, the concept of dimensional reduction follows the Appelquist-Carazzone decoupling theorem \cite{Appelquist:1981vg}. In this idea, the \textit{static mode} is related to a dimensional reduction and corresponds to the limit of high temperatures (small length). 

Concerning the renormalon problem, some authors have explicitly shown that the structure of the renormalon singularities depends on the spatial compactification~\cite{Loewe:1999kw,Cavalcanti:2018thz,Ishikawa:2019tnw,Ashie:2020bvw}. There are indications that new renormalon poles appear once we consider a spatial restriction (or finite temperature) and that all renormalon poles disappear in the limit of dimensional reduction ($L\rightarrow0$). 

Regarding the adiabatic conjecture and the investigation of nonperturbative phenomena, some works consider more than one compactification and also twisted boundary conditions \cite{Anber:2015wha, Anber:2012ig,Simic:2010sv,Sulejmanpasic:2016llc,	Cherman:2016vpt, Buividovich:2017jea, Hongo:2018rpy}. However, there is still a lack of investigation on the behavior of the renormalon singularities when the theory has more than one spatial restriction and examines the role of boundary conditions. Regarding dimensional reduction, in the sense of a $L\rightarrow0$ limit, at the one-loop level we know that the choice of boundary condition plays a vital role ~\cite{Cavalcanti:2018thz,Cavalcanti:2019mli} both for bosonic and fermionic models. This seems to indicate that the insertion of more compactification or boundary conditions might play some role.

The original objective of our study was to fill a small gap and probe the dependence of the renormalon phenomena for more than one spatial restriction and boundary conditions.  However, we found an unexpected result that seems to contradict previous findings in \cite{Loewe:1999kw,Cavalcanti:2018thz,Ishikawa:2019tnw,Ashie:2019cmy,Ashie:2020bvw} and show that the renormalon singularities are \textit{allways} present, even in the limit of dimensional reduction. The reason has its roots in a discussion by Landsman about the limitations on dimensional reduction (see \cite{Landsman:1989be}). It happens because the limit of high temperature / small length does not correspond with the decoupling of static and dynamic modes if we deal with Feynman diagrams of high orders. The reason is that the static mode and the dynamic modes combine to give the final contribution and, therefore, the decoupling theorem fails in this scenario. It means that works that assume a static mode approximation (full - in the sense that all computations consider just the static mode; or partial - in the sense that a part of computation uses the dynamic modes but the remaining consider the static mode) might hide the correct behavior.

In this work, as our interest is rather formal, we consider as a toy model a massless scalar field with quartic self-interaction that carry the relevant structure of the problem. We investigate its behavior regarding the number of spatial compactification, the length scales, and the choice of boundary condition. We show that the renormalon poles \textit{do not disappear} nor the number of singularities changes when we approach a dimensional reduction in the sense of an explicit reduction of the size $L$ of spatial compactification. Regarding the boundary condition, we assume here a twisted boundary condition controlled by some parameter $\theta$ and show that it does not interfere with the number nor location of the poles, only with its residues. For completeness, we also show that if we enforce the decoupling theorem and assume a purely static mode we obtain the ``expected" (based on some previous works) result that the renormalon poles disappear. Therefore, we claim that for our model the renormalon poles disappear not for a smooth dimensional reduction ($L\rightarrow0$) but instead for a static mode approximation.

In section~\ref{sec:renormalons}, we do a quick review of the renormalon phenomena to understand that we need the asymptotic behavior of what we call the chain-link. In section~\ref{sec:compactbubble}, we consider the bubble chain-link (that produces a chain of bubbles) with spatial restrictions and explain the different scales of interest. After that, we follow the scenario with just one compactification as an explicative scenario and investigate the renormalon singularities, sec.~\ref{sec:OneCompactified}. Finally, we extend the treatment to more compactifications and show the effects of the number of compactifications on the renormalon poles, sec.~\ref{sec:MoreCompactifications}. In section~\ref{sec:conclusions}, we expose the conclusions.

\section{Renormalons - a quick review}
\label{sec:renormalons}

Let us consider a renormalizable quantum field theory. We can express some observable $A$ as a perturbative series in the coupling constant $g$,
\begin{equation}
A(g) = \sum_{n=0}^\infty a_n g^n.
\end{equation} 
\noindent At each order in the expansion, we sum up the relevant Feynman diagrams to obtain the coefficient $a_n$. Although the coefficients are finite -- as we deal with a renormalized model -- they behave in many scenarios as a factorial growth (for example $a_n \sim c^n n!$ for large values of $n$, with $c \in \mathbb{R}$). Therefore, if one tries to sum up the perturbative series naively, the value of the observable blows up because the series is divergent. Which is evident in our example,
\begin{equation}
A(g) = \sum_{n=0}^\infty n! (cg)^n.
\label{eq:renorm:an}
\end{equation}

To make sense out of perturbative quantum field theory, one must understand the series as asymptotic (we employ the usual notation $a \sim b$ to say that $a$ is asymptotic to $b$). The perturbative series is just a formal representation of the observable, and one must employ some summation technique to recover the exact nonperturbative expression. One such procedure that is well understood and widely applied is to consider a \textit{Borel summation}. The Borel transform of the series adds a controlling factorial and transport the function from the physical $g$-plane to the Borel $u$-plane,
\begin{equation}
A(g) \sim \sum_{n=0}^\infty a_n g^n \Rightarrow B_A(u) \sim \sum_{n=0}^\infty \frac{a_n}{n!} u^n.
\end{equation}

Assuming that the summation in the Borel plane is well defined one then employs a inverse Borel transform to get back to the usual coupling constant plane $g$,
\begin{equation}
\widetilde{A}(g) = \int_0^\infty du\; e^{-u}B_A(gu).
\label{eq:renorm:invborel}
\end{equation}

\noindent Then, one can relate the original observable $A(g)$ with the exact expression $\widetilde{A}(g)$, that is, $A(g) \sim \widetilde{A}(g)$, meaning that they are asymptotically equivalent.

However, there are many cases where the inverse Borel transform, eq.~\eqref{eq:renorm:invborel}, is not achievable due to singularities in Borel plane that add ambiguities (due to the choice of contour integration). For example, if we take eq.~\eqref{eq:renorm:an},
\begin{equation}
\widetilde{A}(g) = \int_0^\infty du\;  \frac{e^{-u} }{1-cgu},
\end{equation}
\noindent there is a pole located at $gu = 1/c$. The coupling constant is positive, therefore, if $c>0$ the pole lies in integration path (the positive real axis) and disturbs the inverse Borel transform. Integrating over it introduces ambiguities due to the choice of the integration path.

There are some possible sources of singularity in the Borel plane. The proliferation of Feynman diagrams as the perturbative order increases introduces a known factorial growth whose ambiguity is canceled by the semi-classical contribution of instanton configurations by the Bogolmony--Zinn-Justin mechanism~\cite{Bogomolny:1980ur, ZinnJustin:1981dx,Jentschura:2004jg,ZinnJustin:2004cg,Lipatov:1976ny}; due to this relationship, we usually refer to this as ``instanton singularity". The renormalon problem, our main interest, has its roots in the infrared (IR) or ultraviolet (UV) behavior of the internal momentum under integration in a set of Feynman diagrams. A theory can possess both instantons and renormalons singularities. Indeed, it is already well established for QCD that the renormalon singularity controls the UV behavior of the theory as it lies closer to the origin in the Borel plane than the instanton singularity~\cite{Beneke:1998ui}.

It is common, when taking into account renormalon singularities, to assume a bosonic/fermionic field with $N$ flavors and employ a large $N$ approximation. In this approximation, the instanton singularities became far away from the origin and can be ignored (in general, while the instanton poles occur at $\kappa_0 j$ for $j\in \mathbb{N}$, the renormalon singularities occur at $\kappa_0 j/N$). Also, the relevant Feynman diagram is planar in the large-$N$ approximation, and we can already obtain renormalon singularities from it. However, in this article, we \textit{do not} employ a large $N$ approximation (although we could). Instead, we stick with $N=1$ and consider the sum of a set of diagrams that shall at least appear. 

The most common source of renormalons is the chain of bubbles (the usual Adler function, for QCD). Due to historical reasons and extensive usage, it is common to associate renormalons with a chain of bubbles diagram (bubble-chain). However, they are not intrinsically restricted to this~\footnote{In another article, to appear, we discuss renormalon singularities that cames from different sets of Feynman diagrams.}.

As we are interested in a formal description of renormalons within theories with some spatial restriction, it suffices to consider as a toy model a scalar field theory. It carries the relevant structure for a formal investigation on renormalons and its possible dependencies with compactified dimensions and boundary conditions.

\begin{figure}
	\centering
	\includegraphics[width=0.7\linewidth]{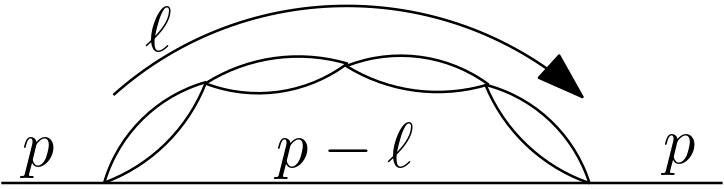}
	\caption{Bubble-chain diagram with 4 bubbles.}
	\label{fig:bubblechain}
\end{figure}
\begin{figure}
	\centering
	\includegraphics[width=0.5\linewidth]{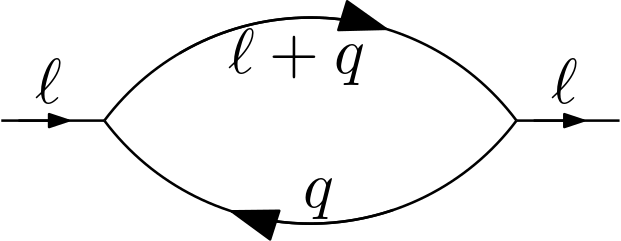}	\caption{Bubble-link diagram.}
	\label{fig:bubble}
\end{figure}

Let us just consider one-chain diagrams whose chain insertions (chain-links) are bubble diagrams (see fig.~\ref{fig:bubblechain}). We also call the chain-links as bubble-links (see fig.~\ref{fig:bubble}), whose amplitude is given by
\begin{align}
\mathcal{B}^D_0(\ell) &= \int \frac{d^D q}{(2\pi)^D} \frac{1}{q^2 + M^2} \frac{1}{(q+\ell)^2 + M^2}\nonumber\\
&= \frac{\Gamma\left[2-\frac{D}{2}\right]}{(4\pi)^\frac{D}{2}} \int_0^1 dz\; \left(M^2 + \ell^2 z(1-z)\right)^{\frac{D-4}{2}}
, \label{eq:renormalon:bubble}
\end{align}
\noindent where $D$ is the number of dimensions, $M$ the mass, $\ell$ the momentum that flows through the chain and $z$ the Feynman parameter introduced to deal with the integration, $\Gamma$ is the usual gamma function. To obtain the dominant behavior of this amplitude for large $\ell$ (UV regime) we rewrite it\footnote{This manipulation is to make evident the presence of the mass. Without it one must be cautious about illusory IR divergence that might appear and add a IR cutoff in the momentum integral. Due to the many manipulations of our treatment, we choose this path to avoid misleading computations.} as
\begin{align}
\mathcal{B}^D_0(\ell) &= \frac{\Gamma\left[2-\frac{D}{2}\right]}{(4\pi)^\frac{D}{2}} (\ell^2+M^2)^{\frac{D-4}{2}}\times\nonumber\\& \int_0^1 dz\;\left(\frac{M^2}{M^2+\ell^2} + \frac{\ell^2}{M^2+\ell^2} z(1-z)\right)^{\frac{D-4}{2}}.
\end{align}
\noindent And we can study its asymptotic behavior\footnote{For $D=4$ and $D=3$ we can take $\frac{M^2}{M^2+\ell^2} + \frac{\ell^2}{M^2+\ell^2} z(1-z) \sim  z(1-z)$. For $D\le2$ we must take $\frac{M^2}{M^2+\ell^2} + \frac{\ell^2}{M^2+\ell^2} z(1-z) \sim \frac{M^2}{M^2+\ell^2} +  z(1-z)$ } for large momentum $\ell$. For $D=4-2\varepsilon$ the bubble amplitude is expected to diverge and we  treat it with the Bogoliubov-Parasiuk-Hepp-
Zimmerman (BPHZ) procedure~\cite{Itzykson:1980rh} to define a finite amplitude, in this scenario this is achieved by the subtraction $\widehat{\mathcal{B}}^4_0(\ell) = \mathcal{B}^4_0(\ell) - \mathcal{B}^4_0(0)$. Then, we obtain 
\begin{equation}
\widehat{\mathcal{B}}^4_0(\ell) \sim - \frac{1}{(4\pi)^2}\ln \frac{\ell^2+M^2}{M^2}.
\label{eq:bulk:regularizedbubble}
\end{equation}

For $D<4$ the amplitude is already finite ($\widehat{\mathcal{B}}^{D<4}_0(\ell) = \mathcal{B}^{D<4}_0(\ell)$) and we have
\begin{equation}
\mathcal{B}_0(\ell) \sim \begin{cases}
\frac{1}{8} \frac{1}{\sqrt{\ell^2+M^2}},& D=3;\\
\frac{1}{2\pi} \frac{\ln \frac{\ell^2+M^2}{M^2} }{\ell^2+M^2} ,& D=2;\\
\frac{1}{M} \frac{1}{\ell^2+M^2} ,& D=1;
\end{cases}
\end{equation}

It is unnecessary to consider $D>4$ as the theory is not renormalizable for dimensions higher than $D=4$. The only scenario where the bulk bubble amplitude produces the logarithmic behavior that can produce renormalons occurs for $D=4$. For odd dimensions, there is no logarithmic behavior, while for other even dimensions, there a damping factor that turns the sum of the bubble chain in the physical plane finite. 

As already commented by some authors~\cite{Beneke:1998ui}, the production of a factorial growth occurs in the perturbative series only if the bubble diagram has a logarithmic behavior with the internal momentum. It means that to occur an IR (UV) renormalon singularity, the chain-link diagram must have a logarithmic behavior for the IR (UV) regimes\footnote{In fact, we can investigate the conditions to be satisfied by the chain-link and the chain structure so that renormalon singularities are present. These are a subject for another article, to appear}. 

The chain with $k$ bubbles is $\mathcal{T}_k(g)$ (see, for example the $k=4$ scenario in fig.~\ref{fig:bubblechain}) and the sum over all chain diagrams gives $\mathcal{T}$, 
\begin{align}
\mathcal{T}(g) \sim&
\sum_{k=0}^\infty \mathcal{T}_k(g)
=
\sum_{k=0}^\infty (- g) \times\nonumber\\&\int \frac{d^D\ell}{(2\pi)^D}\; 
\frac{1}{(p-\ell)^2+M^2} \left[-\frac{g}{2}\widehat{\mathcal{B}}^D_0(\ell)\right]^{k}.
\end{align}

We shall use the naive approximation that $(p-\ell)^2 = p^2+\ell^2$, such that we can integrate over the solid angle and obtain
\begin{align}
\mathcal{T}(g;p,M) \sim &
(- g) \sum_{k=0}^\infty \frac{2}{(4\pi)^\frac{D}{2}\Gamma\left(\frac{D}{2}\right)} \times\nonumber\\&
\int_0^\infty d\ell\; 
\frac{\ell^{D-1}}{P^2+\ell^2} \left[-\frac{g}{2} \widehat{\mathcal{B}}^D_0(\ell)\right]^{k}.
\end{align}
\noindent Each of these amplitudes must yet be regularized and to do so we employ the BPHZ procedure. For $D=4$, as $\widehat{\mathcal{B}}^D_0(\ell) \propto \ln \ell$ it means that we must apply the subtraction
\begin{multline}
\widehat{\mathcal{T}}(g;p,M) = \mathcal{T}(g;p,M) - \mathcal{T}(g;0,M) \\- p^2 \frac{\partial}{\partial p^2} \mathcal{T}(g;p,M) \Bigg|_{p^2=0},
\end{multline}
\noindent in a low-$p$ approximation it produces
\begin{multline}
\widehat{\mathcal{T}}(g;p,M) \sim 
(- g) \sum_{k=0}^\infty \frac{2 p^4}{(4\pi)^2} \int_0^\infty d\ell\; 
\frac{\ell^{3}}{(M^2+\ell^2)^3}\times\\ \left[-\frac{g}{2} \widehat{\mathcal{B}}^4_0(\ell)\right]^{k}.
\end{multline}

The perturbative series defining $\widehat{\mathcal{T}}(g;p,M)$ is asymptotic and expected to diverge with a factorial growth. To investigate its behavior we can transport it from the physical $g$-plane to the Borel $u$-plane\footnote{If one is interested in the BPHZ-regularized expression for $D=4$ without the approximations, it is
	\begin{multline}
	B_{-\mathcal{T}/g}(u) = 2 e^{-\frac{gu}{(4\pi)^2}}  \int\frac{d^D\ell}{(2\pi)^D} \int_0^1 dx (1-x)\times\\
	\frac{p^4+2p\cdot\ell (\ell^2+M^2-p^2)}{\left[(\ell-px)^2+M^2+p^2x(1-x)\right]^3}
	\left[\frac{1+\sqrt{1+4 \frac{M^2}{\ell^2}}}{1-\sqrt{1+4 \frac{M^2}{\ell^2}}}\right]^{\frac{gu}{(4\pi)^2} \sqrt{1+4 \frac{M^2}{\ell^2}}}.
	\end{multline}
}
\begin{align}
B_{-\mathcal{T}/g}(u) &=   \sum_{k=0}^\infty \frac{2 p^4}{(4\pi)^2} \int_0^\infty d\ell\; \ell^{D-1} 
\frac{1}{(M^2+\ell^2)^3} \times\nonumber\\&
\left[-\frac{gu}{2}\mathcal{B}^4_0(\ell)\right]^{k} \frac{1}{k!} \nonumber\\
&=  \frac{2p^4}{(4\pi)^2} \int_0^\infty d\ell\; \ell^{D-1}
\frac{e^{-\frac{gu}{2}\mathcal{B}^4_0(\ell)}}{(M^2+\ell^2)^3},
\end{align}
\noindent note that we follow the convention not to drop $g$ in the Borel plane. In the literature it is common to absorb the $g$ coupling and some multiplicative factor\footnote{Usually the one-instanton action, but it may vary based on convenience}. We shall refer to poles at $\kappa_0 u$, where $\kappa_0= g /\left( 2(4\pi)^2\right)$ is some useful multiplicative constant. Using the known expression for $\widehat{\mathcal{B}}^4_0(\ell)$, eq.~\eqref{eq:bulk:regularizedbubble}, it follows that
\begin{align}
B_{-\mathcal{T}/g}(u) =&\frac{2p^4}{(4\pi)^{2}M^{2\kappa_0 u}} \int_0^\infty d\ell\;
\frac{ \ell^{3}}{(M^2+\ell^2)^{3-\kappa_0 u}} 
\times\nonumber\\&
=  
\frac{p^4}{(4\pi)^{2}M^2} \frac{1}{(2-\kappa_0 u)(1-\kappa_0 u)}.
\end{align}

Meaning that there are two simple poles, located at $\kappa_0 u=1, 2$. A more brute-force procedure (ignoring $M^2$ completely) would give just the first pole\footnote{A partial brute-force, taking for the bubble amplitude $\ln \ell$ but still using $\ell^2+M^2$ to calculate the full $\mathcal{T}$ scenario would lead to
	\begin{equation*}
	\int_0^\infty d\ell\; \frac{\ell^{3+\kappa_0 u}}{(\ell^2+M^2)^3} = \frac{M^{\kappa_0 u-2}}{4} \Gamma(2+\kappa_0 u/2)\Gamma(1-\kappa_0 u/2)
	\end{equation*}
	that have poles for all even integers except $\kappa_0 u = -2,0$. This behavior occurs due to an illusory IR divergence. We must be careful not to fall into this when making asymptotic approximatinos}.
For this class of diagrams, there is no need to look at $D\neq4$ as it does not produce renormalon poles\footnote{For $D>4$, the analysis is meaningless, as the theory is nonrenormalizable and one need to introduce infinite counterterms. One can check that for both $D=3$ and $D=2$ that no factorial growth occurs in the physical $g$-plane.}.

In the following, we take into account the influence of compactified dimensions. We consider that out of the $D$ dimensions of the entire space, $d$ of them have a spatial restriction with some periodic boundary conditions. The compactification length is $L_\alpha$, where $\alpha$ runs from $1$ to $d$ (although we could also make it run fro $0$ to $d-1$ and assume that $L_0 = \beta = 1/T$ is the inverse temperature). The twist angle $\theta$ controls the periodicity, $\theta$ runs from $0$ (periodic boundary condition) to $1$ (antiperiodic boundary condition).  

The basic structure determining the presence of renormalons is the asymptotic behavior of the bubble link. In sec.~\ref{sec:compactbubble} we show the general expression for $d$ compactifications. We first develop all the problems for the scenario with just one compactification in sec.~\ref{sec:OneCompactified}, discussing the behavior of the renormalon for one length scale ($L$) and the twisted boundary condition. After that, we extend it to the scenario with $d$ compactifications and discuss the behavior of the renormalon poles. For completeness, we make a quick discussion on a full static mode approximation in sec.~\ref{sec:StaticMode}.


\section{Compactified Bubble-link}
\label{sec:compactbubble}

Our primary interest is how the dimensionality and size of the system affect a model with renormalons. To tackle it, we investigate the asymptotic behavior of the bubble-link diagram with respect to the external momentum $\ell$ (external to the bubble, internal to the bubble chain) for different compactification lengths $L_i$. We consider some length-scale $\Lambda^{-1}$ so that we can picture different scenarios under interest:
\begin{itemize}
	\item $L_i \gg \Lambda^{-1}, \forall i \in [1,d]$; means that all compactification lengths are dominant with respect to the scale. Therefore, we must be close to a bulk scenario (no size restrictions, $L_i\rightarrow\infty$). 
	\item $\Lambda^{-1}\ll L_{i'} \ll L_{i''}$ for $i'$ running over $d'$ compactified dimensions and $i''$ running over the remaining $d''=d-d'$; in this case there is a somewhat scale split, where $d'$ dimensions are subdominant with respect to the others $d'' = d-d'$ dimensions. However, we still have all compactified dimensions bigger than the length-scale.
	\item $L_{i'} \ll \Lambda^{-1} \ll L_{i''}$ where $i'$ and $i''$ runs s the former; here we also have a split but now we would expect a partial dimensional reduction as some compactified length are subdominant with respect to our scale.
	\item $L_i \ll \Lambda^{-1}, \forall i \in [1,d]$; meaning that all compactification lengths are subdominant compared with the length scale. In this scenario, we might expect that the system is near a dimensional reduction.
\end{itemize}

One must recall that a \textit{dimensional reduction} must be understood as an \textit{effective} behavior. We must understand it in the sense that the ``extra" dimension has a subdominant (or irrelevant) influence. There is no necessity that the characteristic length of the reduced dimension vanishes~\cite{Fisher:1973jvst}. 

In what follows, we stop to mention $\Lambda$ and instead take $\ell L$, where $\ell$ is the momentum external to the bubble diagram and internal to the bubble-chain diagram. Moreover, although we can think about many different scenarios based on the scales of $L_i$ we find in sec.~\ref{sec:OneCompactified} (for $d=1$) and also in sec.~\ref{sec:MoreCompactifications} (for any $d$) that in our model we can achieve an expression valid for all range of $L_i$. Meaning that there is no need to subdivide the analysis into many different regimes.

Let us take the bubble function in a Euclidean space with $D$ dimensions, eq.~\eqref{eq:renormalon:bubble}, and consider that $d\le D$ dimensions are periodically compactified such that we can employ an extension of the imaginary time Matsubara formalism to deal with it~\cite{Khanna:2014qqa}. Each momentum integration is replaced by an infinite sum and the momenta related to the compactified dimensions are replaced by the extended Matsubara frequencies,
\begin{align*}
\int dq &\rightarrow \frac{1}{L} \sum_{n \in \mathbb{Z}},\\
q^2 &\rightarrow \omega_n^2 = \left(\frac{\pi (2n+\theta)}{L} \right)^2,
\end{align*}
\noindent where $\theta$ is the twist angle, that deals with the periodicity of the boundary condition. In our scenario, with a flavorless scalar field, only $\theta=0,1$ are allowed. However, we keep $\theta \in [0,1]$ so that we can adapt the computation for other scenarios in future works if needed. 

After applying the extended Matsubara identity we obtain the bubble function in periodically compactified spaces as
\begin{align}
\mathcal{B}^{D}_{d}(\ell;\omega^\ell_{n_\alpha}) =& \frac{1}{\prod_{\alpha=1}^d L_\alpha} 
\sum_{ \substack{ n_\alpha^q \in \mathbb{Z} \\ \forall \alpha \in [1,d]}} 
\int \frac{d^{D-d} q}{(2\pi)^{D-d}} 
\times\nonumber\\&
\frac{1}{q^2 + \sum_{\alpha=1}^{d} (\omega^q_{n_\alpha})^2 + M^2}\times\nonumber\\& \frac{1}{(q+\ell)^2 + \sum_{\alpha=1}^{d} (\omega^q_{n_\alpha}+\omega^\ell_{n_\alpha})^2 + M^2}
\end{align}
\noindent where the momenta of the compactified dimensions became discrete frequencies $\omega^k_{n_\alpha}$, the superindex identifies the $(D-d)$-momenta related to it. The frequencies are given by
\begin{equation}
\omega^k_{n_\alpha} = \frac{2\pi n^k_\alpha}{L_\alpha} + \frac{\pi \theta_\alpha}{L_\alpha}, n^k_\alpha \in \mathbb{Z}.
\end{equation}

\noindent We consider just lengths, but we could also assume $L_0 = \beta = 1/T$ for a thermal model.

To evaluate the integral over the momenta we introduce the Feynman parametrization that join the propagators with the cost of a new integration over the Feynman parameter $z$, 
\begin{multline}
\mathcal{B}^{D}_{d}(\ell;\omega^\ell_{n_\alpha}) =
\frac{1}{\prod_{\alpha=1}^d L_\alpha} 
\sum_{ \substack{ n_\alpha^q \in \mathbb{Z} \\ \forall \alpha \in [1,d]}} 
\int \frac{d^{D-d} q}{(2\pi)^{D-d}} 
\times\\
\int_{0}^{1}dz \Bigg[(q+\ell z)^2 + \sum_{\alpha=1}^{d} (\omega^q_{n_\alpha}+\omega^\ell_{n_\alpha}z)^2 +\\
+ \left(\ell^2+ \sum_{\alpha=1}^{d}(\omega^\ell_{n_\alpha})^2\right)z(1-z)
+ M^2\Bigg]^{-2},
\end{multline}
\noindent at this point we can make a shift in the internal momentum $q$ as $q\rightarrow q - \ell z$ and absorb it by a translation in the integration sign. However, we cannot do the same with the frequencies $\omega$ because the shift cannot be ``absorbed" with a reparametrization of the sum. The next step is to solve the integral over the momenta by using dimensional regularization techniques,
\begin{multline}
\mathcal{B}^{D}_{d}(\ell;\omega^\ell_{n_\alpha}) =
\frac{1}{\prod_{\alpha=1}^d L_\alpha} 
\sum_{ \substack{ n_\alpha^q \in \mathbb{Z} \\ \forall \alpha \in [1,d]}} 
\frac{\Gamma\left[2 - \frac{D-d}{2}\right]}{(4\pi)^{\frac{D-d}{2}}} \times\\
\int_{0}^{1}dz \Bigg[ \sum_{\alpha=1}^{d} (\omega^q_{n_\alpha}+\omega^\ell_{n_\alpha}z)^2 +\\ + \left(\ell^2+ \sum_{\alpha=1}^{d}(\omega^\ell_{n_\alpha})^2\right)z(1-z) + M^2\Bigg]^{-2+\frac{D-d}{2}}, \label{eq:bubble_provisorio}
\end{multline}
\noindent or, with the modes in evidence,
\begin{multline}
\mathcal{B}^{D}_{d}(\ell;\omega^\ell_{n_\alpha}) =
\frac{1}{\prod_{\alpha=1}^d L_\alpha} 
\sum_{ \substack{ n_i^q \in \mathbb{Z} \\ \forall i \in [1,d]}} 
\frac{\Gamma\left[2 - \frac{D-d}{2}\right]}{(4\pi)^{\frac{D-d}{2}}} 
\int_{0}^{1}dz 
\Bigg[ 
\\
\sum_{i=1}^{d} \frac{4\pi^2}{L_i^2}\left(n^q_i+n^\ell_i+\frac{\theta_i}{2}(1+z)\right)^2+ M^2  \\
+ \left(\ell^2
+ \sum_{i=1}^d \frac{(2\pi n_i^\ell+\pi \theta_i)^2}{L_i^2} \right) z(1-z)
\Bigg]^{-2+\frac{D-d}{2}}.
\label{eq:Bubble_DimRed}
\end{multline}

It is well-known both from finite temperature quantum field theory~\cite{Landsman:1986uw} and QFT with spatial compactifications~\cite{Khanna:2014qqa} that the divergence behavior to be regularized and absorved by renormalization comes only from the bulk (no compactification contribution). Therefore, we can still employ the BPHZ procedure, where the subtractions are done at the bulk ($L_\alpha = \infty, \forall \alpha$). For $D=4$ the expression becames $\mathcal{B}^{D}_{d}(\ell;\omega^\ell_{n_\alpha}) - \mathcal{B}^{D}_{0}(\ell=0)$. 

The sum of all contributions of the Feynman diagrams in the scenario with $d$ compactified dimensions is

\begin{align}
-\frac{\mathcal{T}_k(g)}{g} &= \frac{1}{\prod_{\alpha=1}^d L_\alpha} 
\sum_{\substack{n_\alpha^\ell \in \mathbb{Z}\\ \forall \alpha \in [1,d]}}
\int \frac{d^{D-d}\ell}{(2\pi)^{D-d}}\; 
\times\nonumber\\&
\frac{1}{(p-\ell)^2+(\omega^p_{n_\alpha}-\omega^\ell_{n_\alpha})^2+M^2} \left[-\frac{g}{2}\widehat{\mathcal{B}}^D_d(\boldsymbol{\ell})\right]^{k},
\end{align}
\noindent here we use a simplified notation where
\begin{equation}
\boldsymbol{\ell}^2 = \ell^2+ \boldsymbol{\omega}_\ell^2 = \ell^2 + \sum_{\alpha=1}^d (\omega_{n_\alpha}^\ell)^2.
\end{equation}

In what follows we consider the naive approximation $(a-b)^2 = a^2+b^2$ so that
\begin{align}
-\frac{\mathcal{T}_k(g)}{g}  =& \frac{1}{\prod_{\alpha=1}^d L_\alpha} \sum_{\substack{n_\alpha^\ell \in \mathbb{Z}\\ \forall \alpha \in [1,d]}}
\int \frac{d^{D-d}\ell}{(2\pi)^{D-d}}\; 
\frac{1}{P^2+M^2+ \boldsymbol{\ell}^2}
\times\nonumber\\&
\left[
-\frac{g}{2}\widehat{\mathcal{B}}(\boldsymbol{\ell})
\right]^{k},
\end{align}
\noindent here we define $P^2 = p^2+(\omega^p_{n_\alpha})^2$. Integrating over the solid angle we obtain it becames
\begin{align}
-\frac{\mathcal{T}_k(g)}{g}  =& \frac{1}{\prod_{\alpha=1}^d L_\alpha} \frac{2}{(4\pi)^{\frac{D-d}{2}}\Gamma\left(\frac{D-d}{2}\right)}
\sum_{\substack{n_\alpha \in \mathbb{Z}\\ \forall \alpha \in [1,d]}}
\times\nonumber\\&
\int_0^\infty d\ell\; 
\frac{\ell^{D-d-1}}{P^2+M^2+ \boldsymbol{\ell}^2} \left[
-\frac{g}{2}\widehat{\mathcal{B}}(\boldsymbol{\ell})
\right]^{k}.
\end{align}
\noindent Just like the scenario with no compactifications (see Sec.~\ref{sec:renormalons}) we have to employ the BPHZ procedure at $D=4$ because the amplitude is divergent. For low-$P$ it produces the modification
\begin{equation*}
\frac{1}{P^2+M^2+\ell^2} \rightarrow \frac{P^4}{(\ell^2+M^2)^3},
\end{equation*}
\noindent and the $k$-bubble chain becames
\begin{align}
-\frac{\widehat{\mathcal{T}}_k(g)}{g}  =& \frac{1}{\prod_{\alpha=1}^d L_\alpha} \frac{2P^4}{(4\pi)^{\frac{D-d}{2}}\Gamma\left(\frac{D-d}{2}\right)}
\sum_{\substack{n_\alpha^\ell \in \mathbb{Z}\\ \forall \alpha \in [1,d]}}
\nonumber\\&
\int_0^\infty d\ell\; 
\frac{\ell^{D-d-1}}{(M^2+ \boldsymbol{\ell}^2)^3} \left[
-\frac{g}{2}\widehat{\mathcal{B}}(\boldsymbol{\ell})
\right]^{k}.
\label{eq:bubblechain_regularized}
\end{align}
The sum over the bubble-chain diagrams $\mathcal{T}_k$ is expected to be divergent. To investigate the singularities we change from the physical $g$-plane to the Borel $u$-plane. And, just like sec.~\ref{sec:renormalons} we keep the factor $g$ explictly,
\begin{align}
B_{-\widehat{\mathcal{T}}/g}(u) =  & \frac{1}{\prod_{\alpha=1}^d L_\alpha} \frac{2P^4}{(4\pi)^{\frac{D-d}{2}}\Gamma\left(\frac{D-d}{2}\right)}
\sum_{\substack{n_\alpha^\ell \in \mathbb{Z}\\ \forall \alpha \in [1,d]}}
\nonumber\\&
\int_0^\infty d\ell\; 
\frac{\ell^{D-d-1}}{(M^2+ \boldsymbol{\ell}^2)^3}
e^{-\frac{gu}{2}\widehat{\mathcal{B}}^D_d(\boldsymbol{\ell})}
\label{eq:bubblechain_regularized_borel}
\end{align}

The exact expression for $\mathcal{B}^D_d(\boldsymbol{\ell})$ might depend on the specific scenario under consideration. We find, however, that our expression has a unique asymptotic expression for large $\ell$ that is valid for all range of $L_i$.  We first consider in sec.~\ref{sec:OneCompactified} the scenario with just one compactified dimension so that we can investigate more clearly the behavior. With this, we obtain the location and residues of the renormalon singularities. After that we generalize the result to the scenario with $d$ compactified dimensions in sec.~\ref{sec:MoreCompactifications}.

\section{Renormalon Poles : one compactified dimension}
\label{sec:OneCompactified}

In the following we mantain $D=4$ fixed and consider the scenario with just one compactified dimension ($d=1$). These choices we expect that the approximations and procedures we use became more clear. The bubble-link contribution, Eq.~\eqref{eq:Bubble_DimRed}, in this scenario is
\begin{align}
\mathcal{B}_1^4 =& \frac{1}{8\pi} \sum_{n^q\in \mathbb{Z}} \int_0^1 dz
\Bigg\{
(ML)^2 
\nonumber\\&
+ \left[(L\ell)^2 
+ (2\pi n^\ell+\pi \theta)^2\right] z(1-z)
\nonumber\\&
+ 4\pi^2 \left[n^q  +n^\ell + \frac{\theta}{2}(1+z)\right]
\Bigg\}^{-\frac{1}{2}}
.
\end{align} 
\noindent The sum over the frequencies $n^q$ can be computed employing the usual representation of the Elizalde zeta function~\cite{Elizalde:2012zza}, in this scenario it gives
\begin{multline}
\sum_{n^q\in \mathbb{Z}} \frac{1}{\sqrt{(n^q+a)^2+\delta}}
=
\frac{\Gamma(\varepsilon)}{\delta^\varepsilon}
\\+ 4 \sum_{n^q\in \mathbb{N}^*} \cos(2\pi n^q a) K_0\left(2\pi n^q \sqrt{\delta}\right).
\end{multline}
\noindent Therefore we obtain for the bubble-link function
\begin{multline}
\mathcal{B}_1^4 = \frac{\Gamma(\varepsilon)}{16\pi^2} 
\int_0^1 dz\; \Bigg[\left(\frac{ML}{2\pi}\right)^2
+ \left(\left(\frac{L\ell}{2\pi}\right)^2+\left(n^\ell+\frac{\theta}{2}\right)^2\right)\times\\z(1-z)
\Bigg]^{-\varepsilon}
+\frac{1}{4\pi^2} \sum_{n^q\in \mathbb{N}^*}
\int_0^1 dz\;
\cos\left[2\pi n^q \left(n^\ell z + \frac{\theta}{2}(1+z)\right)\right]\\ \times
K_0 \left[n^q \sqrt{(ML)^2 + \left[(L\ell)^2+(2\pi n^\ell+\pi \theta)^2\right]z(1-z)}\right],
\label{eq:bcalD4d1_begin}
\end{multline}
\noindent where $K_\nu(z)$ is the modified Bessel function of the second kind (K-Bessel function). We need to renormalize this expression. We deal with it using the BPHZ procedure. As commented before, all divergence occurs in the bulk (infinite size) contribution. Applyingthe usual BPHZ procedure to the $L=0$ component, it provides
\begin{equation*}
-\frac{\Gamma(\varepsilon)}{16\pi^2} 
\left(\frac{ML}{2\pi}\right)^{-2\varepsilon}.
\end{equation*}

Let us deal with the first component of the bubble link $\mathcal{B}^4_1$ adding the subtraction term to it. We can expand for small $\varepsilon$ and obtain
\begin{widetext}
\begin{align}
&\frac{\Gamma(\varepsilon)}{16\pi^2} 
\int_0^1 dz\; \left[\left(\frac{ML}{2\pi}\right)^2
+ \left[\left(\frac{L\ell}{2\pi}\right)^2+\left(n^\ell+\frac{\theta}{2}\right)^2\right]z(1-z)
\right]^{-\varepsilon}
-\frac{\Gamma(\varepsilon)}{16\pi^2} 
\left(\frac{ML}{2\pi}\right)^{-2\varepsilon}\nonumber\\
=&
-\frac{1}{16\pi^2} 
\int_0^1 dz\;
\ln \left[
1
+ \left[\left(\frac{\ell}{M}\right)^2+\left(\frac{2\pi}{ML}\right)^2\left(n^\ell+\frac{\theta}{2}\right)^2\right]z(1-z)
\right].
\end{align}
	\end{widetext}
This integral can be done explicitly. Let us consider $b=\left[\left(\frac{\ell}{M}\right)^2+\left(\frac{2\pi}{ML}\right)^2\left(n^\ell+\frac{\theta}{2}\right)^2\right]$, such that
\begin{equation}
\int_0^1 dz\; \ln \left(1+bz(1-z)\right) = -2 - \sqrt{\frac{b+4}{b}} \ln \frac{\sqrt{b+4}-\sqrt{b}}{\sqrt{b+4}+\sqrt{b}}.
\end{equation}

Now we must investigate the different scenarios under interest. In the bulk ($L\rightarrow\infty$) the variable $b$ behaves as $b \rightarrow (\ell/M)^2 \gg 1$ which allows ourselves to consider a large $b$ expansion. The limit of dimensional reduction ($L\rightarrow 0$), on the other hand, can be taken as $\ell L \ll 1$ ($\ell \rightarrow \infty$ but $L \rightarrow 0$ in such a way that guarantees $\ell L \ll 1$), this means that $b$ behaves as 
\begin{equation*}
b = \frac{1}{(ML)^2}\left[(\ell L)^2 + (2\pi n^\ell + \pi \theta)^2
\right] \sim \left(\frac{2\pi}{ML}\left(n^\ell+\frac{\theta}{2}\right)\right)^2.
\end{equation*}
\noindent As $L\rightarrow0$ we clearly have $b\gg1$, which allows again a lage $b$ expansion. Therefore, that the large $b$ expansion is valid for all range of the compactification parameter $L$. The exception occur only at $\theta=n^\ell=0$. However, under the bubble chain we must sum $n^\ell$ over all integers. Although the static mode $n^\ell=0$ represents an exception there is no motive to ignore the dynamical modes and they represent a dominant contribution. For large $b$ we get
\begin{equation}
\int_0^1 dz\; \ln \left(1+bz(1-z)\right) \sim -2 + \ln b,
\end{equation}
\noindent wich means, for any value of $L$ that the first term of eq.~\eqref{eq:bcalD4d1_begin} is asymptotically equal to
\begin{equation}
\frac{1}{8\pi^2} - \frac{1}{16\pi^2} \ln \left(\frac{\ell^2}{M^2} + \frac{(2\pi n^\ell + \pi \theta)^2}{M^2L^2}\right)
\label{eq:onecomp:static}
\end{equation}

The second term of eq.~\eqref{eq:bcalD4d1_begin} involves a summation over the $n^q$ modes, a sum of K-Bessel functions,
\begin{align}
\frac{1}{4\pi^2} \sum_{n^q\in \mathbb{N}^*}
\int_0^1 dz\;
\cos\left[2\pi n^q \left(n^\ell z + \frac{\theta}{2}(1+z)\right)\right]
\times \nonumber\\K_0 \left[n^q \sqrt{(ML)^2 + \left[(L\ell)^2+(2\pi n^\ell+\pi \theta)^2\right]z(1-z)}\right].
\label{eq:bcalD4d1_Kbessel}
\end{align}
\noindent Let us analyze the behavior for different scales of $L$. Near the bulk ($L\rightarrow\infty$) the dominant behavior is $n^q L\sqrt{M^2+\ell^2 z(1-z)} \gg 1$, meaning that the argument of the K-Bessel is large. Near the dimensional reduction ($L\rightarrow0$) the dominant behavior is $n^q (2\pi n^\ell+\pi \theta)\sqrt{z(1-z)}$, we can see that the argument goes to zero only if $n^\ell=\theta=0$. Otherwise, we can assume that the argument of the K-Bessel function is positive and larger than 1 also near the dimensional reduction. For completeness we could split it into the dynamic modes ($n^\ell\in\mathbb{Z}^*$) that follows perfectly this argument and the static mode in periodic boundary conditions (the exception $n^\ell=\theta=0$). Sticking with the dynamic modes (\textit{or if we consider antiperiodic boundary conditions}) we see that for all range of the compactification length $L$ the argument of the K-Bessel function can be taken as large. We can compare it with the investigation done in Ref.~\cite{Ashie:2020bvw}, there the argument of the K-Bessel function goes to zero in the limit of $L\rightarrow 0$ and an expansion for small values of the parameter is needed; in this scenario the K-Bessel function cancels out the logarithm behavior and the renormalon poles disappear. This example indicates that we do not have a general rule, independent of the theory, on how the renormalon poles behave with respect to the compactification. If we use the asymptotic expansion of the Bessel function for large arguments the dominant contribution cames from $n^q=1$ and produces
\begin{align}
\frac{1}{4\pi^2} \sqrt{\frac{\pi}{2}}
\int_0^1 dz\;
\cos\left[2\pi n^q \left(n^\ell z + \frac{\theta}{2}(1+z)\right)\right]\times \nonumber\\
\frac{e^{\sqrt{(ML)^2 + \left[(L\ell)^2+(2\pi n^\ell+\pi \theta)^2\right]z(1-z)}}}{\left[(ML)^2 + \left[(L\ell)^2+(2\pi n^\ell+\pi \theta)^2\right]z(1-z)\right]^{\frac{1}{4}}}.
\end{align}
\noindent To evaluate this integral we split the region of integration between two similar sectors $z\in[0,1/2]$ e $z \in [1/2,1]$ and apply the change of variables $\phi = \sqrt{(ML)^2 + \left[(L\ell)^2 + (2\pi n^\ell+\pi \theta)^2\right]z(1-z)}$. We obtain that
\begin{widetext}
\begin{align*}
d\phi = \frac{(L\ell)^2+(2\pi n^\ell+\pi \theta)^2}{2\phi} (1-2z)dz\\
1-2z = f_1, \quad z\in[0,1/2];\\
1-2z = -f_1, \quad z\in[1/2,1];\\
f_1^2 = 1 + \frac{(2ML)^2}{(L\ell)^2+(2\pi n^\ell+\pi \theta)^2} - \frac{(2\phi)^2}{(L\ell)^2+(2\pi n^\ell+\pi \theta)^2}.
\end{align*}
The change of variables produces,
\begin{multline}
\frac{1}{4\pi^2} \sqrt{\frac{\pi}{2}}
\int_{ML}^{\sqrt{(ML)^2+\frac{(L\ell)^2+(2\pi n^\ell+\pi \theta)^2}{4}}}d\phi\;
\frac{2\phi}{(L\ell)^2+(2\pi n^\ell+\pi \theta)^2}
\frac{e^{-\phi}\phi^{-\frac{1}{2}}}{\sqrt{
1 + \frac{(2ML)^2}{(L\ell)^2+(2\pi n^\ell+\pi \theta)^2} - \frac{(2\phi)^2}{(L\ell)^2+(2\pi n^\ell+\pi \theta)^2}}}\times \\
\Bigg\{
\cos\left[\pi \theta 
+ \pi \left(n^\ell +\frac{\theta}{2}\right)
\left(1-\sqrt{
	1 + \frac{(2ML)^2}{(L\ell)^2+(2\pi n^\ell+\pi \theta)^2} - \frac{(2\phi)^2}{(L\ell)^2+(2\pi n^\ell+\pi \theta)^2}}\right)
\right]\\
+
\cos\left[\pi \theta 
+ \pi \left(n^\ell +\frac{\theta}{2}\right)
\left(1+\sqrt{
	1 + \frac{(2ML)^2}{(L\ell)^2+(2\pi n^\ell+\pi \theta)^2} - \frac{(2\phi)^2}{(L\ell)^2+(2\pi n^\ell+\pi \theta)^2}}\right)
\right]
\Bigg\},
\end{multline}
\end{widetext}
\noindent and once again we can investigate the dominant behavior by looking at the parameter $b = (L\ell)^2+(2\pi n^\ell+\pi \theta)^2$. As discussed before, $b$ is allways large for all range of $L$, meaning that the dominant contribution from that integral is
\begin{align}
&\frac{\sqrt{2\pi}}{4\pi^2} 
\int_{ML}^{\infty}d\phi\;
\frac{e^{-\phi}\phi^{\frac{1}{2}}}{(L\ell)^2+(2\pi n^\ell+\pi \theta)^2}\times \nonumber\\
&\Bigg[
\cos\left(\pi \theta \right)
+
\cos\left(2\pi \theta + 2\pi n^\ell \right)
\Bigg]\nonumber 
\\=&
\frac{\sqrt{2\pi}}{4\pi^2} 
\frac{\cos\left(\pi \theta \right)
	+
	\cos\left(2\pi \theta \right)
}{(L\ell)^2+(2\pi n^\ell+\pi \theta)^2} \Gamma\left(\frac{3}{2},ML\right).
\label{eq:onecomp:dynamic}
\end{align}

Finally, we can join eq.~\eqref{eq:onecomp:static} and eq.~\eqref{eq:onecomp:dynamic} to obtain the asymptotic behavior of the bubble-link $\widehat{\mathcal{B}}_1^4$ for large $\ell$,
\begin{align}
\widehat{\mathcal{B}}_1^4 \sim
\frac{1}{8\pi^2} - \frac{1}{16\pi^2} \ln \left(\frac{\ell^2}{M^2} + \frac{(2\pi n^\ell + \pi \theta)^2}{M^2L^2}\right)
\nonumber\\+
\frac{\sqrt{2\pi}}{4\pi^2} 
\frac{\cos\left(\pi \theta \right)
	+
	\cos\left(2\pi \theta \right)
}{(L\ell)^2+(2\pi n^\ell+\pi \theta)^2} \Gamma\left(\frac{3}{2},ML\right).
\end{align}

However, as we discussed in the scenario without any compactification, see sec.~\ref{sec:renormalons}, we must be careful with the asymptotic expression under use. One can note that the original expression has a $M^2$ component that act as a natural IR cutoff for small values of $\ell$. If we ignore it completly we might modify the structure of singularities in the Borel plane. To prevent it we modify the above asymptotic expression to include this component,
\begin{align}
\widehat{\mathcal{B}}_1^4 \sim&
\frac{1}{8\pi^2} - \frac{1}{16\pi^2} \ln \left( \frac{(ML)^2 + (\ell L)^2+(2\pi n^\ell + \pi \theta)^2}{M^2L^2}\right)\nonumber
\\&+
\frac{\sqrt{2\pi}}{4\pi^2} 
\frac{\cos\left(\pi \theta \right)
	+
	\cos\left(2\pi \theta \right)
}{(ML)^2+(L\ell)^2+(2\pi n^\ell+\pi \theta)^2} \Gamma\left(\frac{3}{2},ML\right).
\label{eq:bubblelink:d1}
\end{align}


Previous works suggested the disappearance of renormalons singularities when the model under study goes to the limit of dimensional reduction. Indeed, it is possible to achieve this behavior IF we keep only the contribution from the static mode. Due to the Appelquist-Carazzone decoupling theorem~\cite{Appelquist:1981vg}, it is usual to associate a dimensional reduction in a thermal field theory ($T\rightarrow\infty$) with the decoupling of static and dynamic modes at large temperatures. However, as discussed in a seminal paper by Landsman (see ref.~\cite{Landsman:1989be}) ``the dimensional reduction (\textit{in the sense of decoupling of modes}) holds in the lowest order, but breaks down in higher orders of perturbation theory". We take into account the sum over bubble chains, which is a higher-order contribution. Therefore, there is a mixture between the static and dynamic modes, and the decoupling theorem does not hold. In sec.~\ref{sec:StaticMode}, we show for completeness that if we assume the decoupling theorem from the start and employ a static mode approximation, the bubble link behaves in such a way that there is no renormalon singularity. Some works still consider a somewhat equivalence between the static mode approximation and the dimensional reduction or ignore at first approximation the contribution from the dynamic modes. As we discuss here, this kind of approximation might be nocive, in the sense that it might hide the correct behavior. We must also remark that this conclusion is not general, as one can find models where the disappearance still occurs, see ref.~\cite{Ashie:2020bvw}.

Let us now use the asymptotic expression of the bubble link, eq.~\eqref{eq:bubblelink:d1}, to study the structure of singularities of the sum over the set of bubble-chain diagrams. In the Borel plane, see eq.~\eqref{eq:bubblechain_regularized_borel}, we obtain
\begin{multline}
B_{-\frac{\widehat{\mathcal{T}}}{g}}
\sim \frac{P^4 e^{ -\frac{gu}{16\pi^2}} }{2\pi^2 L}
\sum_{n^\ell \in \mathbb{Z}}
\int_0^\infty d\ell
\frac{\ell^2}{(\ell^2+\omega_\ell^2+M^2)^3}
\times \\
\left(\frac{M^2+\ell^2+\omega_\ell^2}{M^2}\right)^{\frac{gu}{32\pi^2}}\times\\
e^{ -\frac{gu\sqrt{2\pi}}{8\pi^2} 
	\frac{\cos\left(\pi \theta \right)
		+
		\cos\left(2\pi \theta + 2\pi n^\ell \right)
	}{(ML)^2+(L\ell)^2+(L\omega_\ell)^2} \Gamma\left(\frac{3}{2},ML\right)
}.
\end{multline}

If we expand the last exponential in power series,
\begin{equation}
e^{-\frac{c}{M^2+\ell^2+\omega_\ell^2}} = \sum_{k \in \mathbb{N}} \frac{1}{k!} \left(-\frac{c}{M^2+\ell^2+\omega_\ell^2}\right)^k,
\end{equation}
\noindent and as $n^\ell \in \mathbb{Z}$, we have $\cos(2\pi \theta+2\pi n^\ell) = \cos(2\pi \theta)$ and get that
\begin{multline}
B_{-\frac{\widehat{\mathcal{T}}}{g}}
\sim \frac{P^4 e^{ -\frac{gu}{16\pi^2}} }{2\pi^2 L M^{\frac{gu}{16\pi^2}}}
\sum_{k \in \mathbb{N}} \frac{1}{k!}
\Bigg[
-\frac{gu\sqrt{2\pi}}{8\pi^2L^2} 
\Gamma\left(\frac{3}{2},ML\right)
\times\\
\left(\cos\left(\pi \theta \right) + \cos\left(2\pi \theta \right)\right)
\Bigg]^k\times\\
\sum_{n^\ell \in \mathbb{Z}}
\int_0^\infty d\ell
\frac{\ell^2}{(\ell^2+\omega_\ell^2+M^2)^{3-\frac{gu}{32\pi^2}+k}}.
\end{multline}

The integral over the momentum $\ell$ is done as
\begin{equation}
\int_0^\infty d{\ell} \frac{{\ell}^{s-1}}{\left({\ell}^2 + x^2\right)^t} = \frac{(x^2)^{\frac{s}{2}-t}}{2} \frac{\Gamma\left(\frac{s}{2}\right)\Gamma\left(t-\frac{s}{2}\right)}{\Gamma(t)},
\end{equation}
\noindent this produces the gamma function. At this point one can notice that the poles of gamma function are responsible for poles of the function $B_{-\frac{\widehat{\mathcal{T}}}{g}}$. About the approximations, one can note that if the $t$ variable depends on $u$ we get \textit{less} poles than if the $s$ variable depends on $u$. 

\begin{multline}
B_{-\frac{\widehat{\mathcal{T}}}{g}}
\sim \frac{P^4 e^{ -\frac{gu}{16\pi^2}} }{4\pi^2 L M^{\frac{gu}{16\pi^2}}}
\sum_{k \in \mathbb{N}} \frac{1}{k!}
\Bigg[
-\frac{gu\sqrt{2\pi}}{8\pi^2L^2} 
\Gamma\left(\frac{3}{2},ML\right) \times\\
\left(\cos\left(\pi \theta \right) + \cos\left(2\pi \theta \right)\right)
\Bigg]^k 
\sum_{n^\ell \in \mathbb{Z}}
\frac{\Gamma\left(\frac{3}{2}\right)\Gamma\left(\frac{3}{2}-\frac{gu}{32\pi^2}+k\right)}{\Gamma\left(3-\frac{gu}{32\pi^2}+k\right)}\times\\
\frac{1}{(\omega_\ell^2+M^2)^{\frac{3}{2}-\frac{gu}{32\pi^2}+k}}.
\end{multline}

The sum over $n^\ell$ is a Elizalde zeta function~\cite{Elizalde:2012zza}, that gives
\begin{multline}
\sum_{n^\ell \in \mathbb{Z}}
\frac{L^{2t}\Gamma(t)}{\left[\left(2\pi n^\ell+\pi \theta\right)^2+(ML)^2\right]^t}
=
\frac{L^{2t}}{2\sqrt{\pi}}
\Bigg\{\\
(ML)^{-2t+1}\Gamma\left(t-\frac{1}{2}\right)
\\+ 4 \sum_{n^\ell\in \mathbb{N}^*} \left(\frac{n^\ell}{2ML}\right)^{t-\frac{1}{2}}
\cos(n^\ell \pi \theta) K_{t-\frac{1}{2}}\left(n^\ell ML\right)
\Bigg\},
\end{multline}
\noindent and the asymptotic approximation for the sum over the bubble-chain diagrams in the Borel plane is
\begin{widetext}
	\begin{align}
	B_{-\frac{\widehat{\mathcal{T}}}{g}}
	\sim &\frac{P^4L^2 e^{ -2\kappa_0 u} }{(4\pi)^2 (ML)^{2\kappa_0 u}}
	\sum_{k \in \mathbb{N}} \frac{1}{k!}
	\left(
	-\kappa_0u\right)^k
	\left(\sqrt{32\pi}\Gamma\left(\frac{3}{2},ML\right)\right)^k 
	\frac{\left(\cos\left(\pi \theta \right) + \cos\left(2\pi \theta \right)\right)^k}{\Gamma\left(3-\kappa_0 u+k\right)}
	\times\nonumber\\&
	\Bigg\{
	(ML)^{-2+2\kappa_0 u-2k}\Gamma\left(1-\kappa_0 u+k\right)
	+ 4 \sum_{n^\ell\in \mathbb{N}^*} \left(\frac{n^\ell}{2ML}\right)^{1-\kappa_0 u+k}
	\cos(n^\ell \pi \theta) 
	K_{1-\kappa_0 u+k}\left(n^\ell ML\right)
	\Bigg\}
	\end{align}
The only singularities in the Borel $u$-plane come from the gamma function. There are a countable infinite set of poles located at the positive integers $\kappa_0 u = j \in \mathbb{N}^*$, with $\kappa_0=g/(2(4\pi)^2)$. It is important to remark that this result is \textit{independent} of the value of $L$. This means that the renormalon singularity is a permanent characteristic of this compactified model.
\end{widetext}

\begin{figure}
	\centering
	\includegraphics[width=\linewidth]{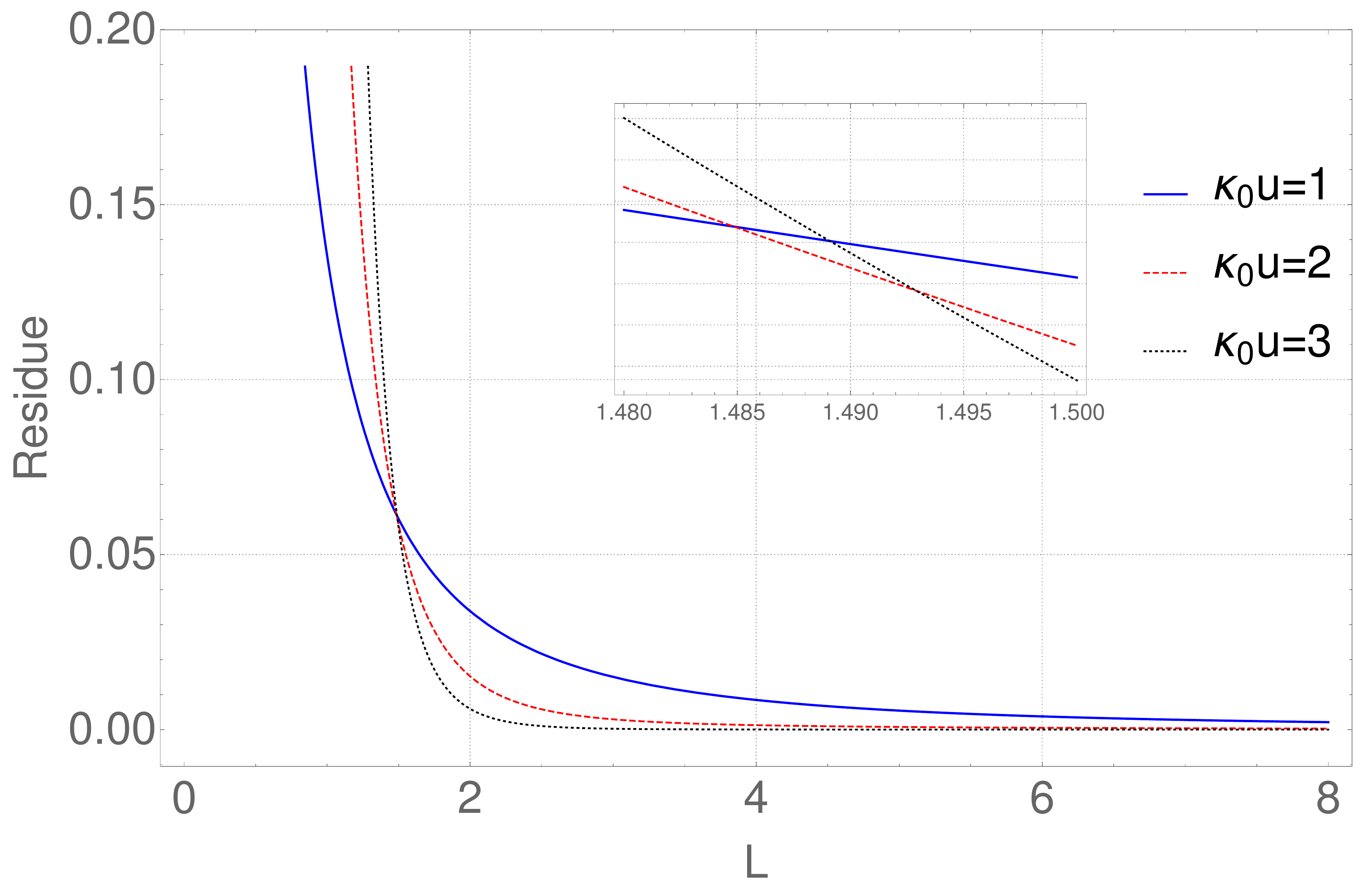}
	\caption{Residues for the first three renormalon poles as a function of the length parameter $L$. This scenario considers $d=1$, $\theta=0$ (periodic boundary condition) and $M=1$. The change in ``hierarchy" occurs around $L=1.490(5)$.}
	\label{fig:residue}
\end{figure}

The location and number of poles are independent of the size of the compactification length $L$. However, the residues of the poles vary with  $L$. We see, for example, a somewhat ``hierarchy". For large values of $L$, the further away the pole is from the origin of the Borel plane, the smaller is the residue. This behavior changes for small values of $L$, where the further away from the origin, the bigger the residue. There is some region at intermediate values of the length $L$ where this ``hierarchy" is changed. We show this behavior at fig.~\ref{fig:residue} for the first three renormalon poles ($\kappa_0 u =1,2,3$).

It is important to remark that brute approximations modify the structure of singularities. For example, if we take the near-bulk limit, one would expect the dominant contribution to be $\widehat{\mathcal{B}}^4_1 \sim \frac{1}{8\pi^2}-\frac{1}{16\pi^2} \ln \frac{\ell^2+M^2}{M^2}$. But then the sum over the bubble-chain contribution, in the Borel plane, becames 
\begin{align}
B_{-\frac{\widehat{\mathcal{T}}}{g}}
\sim& \frac{P^4 e^{ -\frac{gu}{16\pi^2}} }{2\pi^2 L}
\sum_{n^\ell \in \mathbb{Z}}
\int_0^\infty d\ell
\frac{\ell^2}{(\ell^2+\omega_\ell^2+M^2)^3}\times\nonumber\\&
\left(\frac{M^2+\ell^2}{M^2}\right)^{\frac{gu}{32\pi^2}},
\end{align}
\noindent and we get new poles at half-integers values of $\kappa_0 u$. But these half-integer poles seems to be spurious results due to the approximation, as they vanish when we consider the general scenario. This is an indication that the appearence of new renormalon poles must be taken cautiously as the asymptotic approximation under use might hide something.

\section{Renormalon poles : $d$ compactified dimensions}
\label{sec:MoreCompactifications}

At this section we consider the scenario with more compatifications where the bubble function is given by eq.~\eqref{eq:Bubble_DimRed}. Let us first consider the different regimes (scales of length) under interest. To do so we take the compactified lengths as $L_\alpha=L$, with this the bubble link becames
\begin{multline}
\mathcal{B}^{D}_{d}(\ell;\omega^\ell_{n_\alpha}) =
L^{4-D}
\sum_{ \substack{ n_i^q \in \mathbb{Z}^d}} 
\frac{\Gamma\left[2 - \frac{D-d}{2}\right]}{(4\pi)^{\frac{D-d}{2}}} 
\int_{0}^{1}dz 
\Bigg[ 
\\\sum_{i=1}^d
4\pi^2\left(n^q_i+n^\ell_i+\frac{\theta_i}{2}(1+z)\right)^2 
\\
+ (ML)^2
+ \left((\ell L)^2
+ \sum_{i=1}^d (2\pi n_i^\ell+\pi \theta_i)^2 \right) z(1-z)
\Bigg]^{-2+\frac{D-d}{2}}.
\end{multline}
\noindent The scenario near bulk and near dimensional reduction produces simply
\begin{align}
c = (ML)^2
+ \left((\ell L)^2
+ \sum_{i=1}^d (2\pi n_i^\ell+\pi \theta_i)^2 \right) z(1-z)\nonumber\\
=
\begin{cases}
(\ell L)^2z(1-z); &L\rightarrow\infty\\
\sum_{i=1}^d (2\pi n_i^\ell+\pi \theta_i)^2  z(1-z); &L\rightarrow 0
\end{cases}
\end{align}

\begin{widetext}
\noindent meaning that $c$ can always be taken as large, regardless of the length scale. Therefore, we can safely apply the Elizalde zeta analytical extension~\cite{Elizalde:2012zza} that is useful in this regime\footnote{The analytical extension for the Elizalde zeta, \begin{equation}
	\sum_{n_i \in \mathbb{Z}^d} \left(\sum_i a_i(n_i+b_i)^2 + c\right)^{-s} \label{eq:footnote}
	\end{equation} is exact regardless of $c$. However, it is only useful if $c\neq0$. At $c=0$ the infinite sum of K-Bessel functions breaks down and cancels out the `bulk' component.},
\begin{multline}
\mathcal{B}^{D}_{d}(\ell;\omega^\ell_{n_\alpha}) =
\frac{1}{\prod_{\alpha=1}^d L_\alpha} 
\sum_{ \substack{ n_\alpha^q \in \mathbb{Z} \\ \forall \alpha \in [1,d]}} 
\frac{\Gamma\left[2 - \frac{D-d}{2}\right]}{(4\pi)^{\frac{D-d}{2}}} 
\int_{0}^{1}dz 
\Bigg[ 
\sum_{\alpha=1}^{d} \frac{4\pi^2}{L_\alpha^2}\left(n^q_i+n^\ell_i+\frac{\theta_i}{2}(1+z)\right)^2 
+ M^2
\\+ \left(\ell^2
+ \sum_{i=1}^d \frac{(2\pi n_i^\ell+\pi \theta_i)^2}{L_i^2} \right) z(1-z)
\Bigg]^{-2+\frac{D-d}{2}}
= \frac{1}{(4\pi)^{\frac{D}{2}}}\int_0^1 dz\; 
\Bigg\{
\left[M^2+\left(\ell^2+\sum_{i=1}^d \frac{(2\pi n_i^\ell+\pi \theta_i)^2}{L_i^2}\right)z(1-z)\right]^{\frac{D}{2}-2}
\Gamma\left(2-\frac{D}{2}\right)
\\+
4 \sum_{i=1}^d \sum_{n_i \in \mathbb{N}^*}
\left(\frac{n_i L_i}{2 \sqrt{M^2+\left(\ell^2+\sum_{j=1}^d \frac{(2\pi n_j^\ell+\pi \theta_j)^2}{L_j^2}\right)z(1-z)}}\right)^{2-\frac{D}{2}}
 \cos\left(\pi \theta_i(1+z)\right) \times
\\K_{2-\frac{D}{2}}\left(n_i L_i \sqrt{M^2+\left(\ell^2+\sum_{j=1}^d \frac{(2\pi n_j^\ell+\pi \theta_j)^2}{L_j^2}\right)z(1-z)} \right)
\\+
2^{d+1} \sum_{n_1,\ldots,n_d \in \mathbb{N}^*}
\left(\frac{\sqrt{\sum_{i=1}^d n_i^2 L_i^2}}{2 \sqrt{M^2+\left(\ell^2+\sum_{j=1}^d \frac{(2\pi n_j^\ell+\pi \theta_j)^2}{L_j^2}\right)z(1-z)}}\right)^{2-\frac{D}{2}}
\prod_{i=1}^d\cos\left(\pi \theta_i(1+z)\right) \times
\\K_{2-\frac{D}{2}}\left(\sqrt{\sum_{i=1}^d n_i^2 L_i^2} \sqrt{M^2+\left(\ell^2+\sum_{j=1}^d \frac{(2\pi n_j^\ell+\pi \theta_j)^2}{L_j^2}\right)z(1-z)} \right)
+ \ldots
\Bigg\}.
\end{multline}
\noindent At $D=4-2\varepsilon$ this simplifies to
\begin{multline}
\mathcal{B}^{4}_{d}(\ell;\omega^\ell_{n_\alpha})
= \frac{1}{(4\pi)^{2}}\int_0^1 dz\; 
\Bigg\{
\left[M^2+\left(\ell^2+\sum_{i=1}^d \frac{(2\pi n_i^\ell+\pi \theta_i)^2}{L_i^2}\right)z(1-z)\right]^{-\varepsilon}
\Gamma\left(\varepsilon\right)
\\+
4 \sum_{i=1}^d \sum_{n_i \in \mathbb{N}^*}
\cos\left(\pi \theta_i(1+z)\right)
K_{0}\left(n_i L_i \sqrt{M^2+\left(\ell^2+\sum_{j=1}^d \frac{(2\pi n_j^\ell+\pi \theta_j)^2}{L_j^2}\right)z(1-z)} \right)
+ \ldots
\Bigg\}.
\end{multline}
	\end{widetext}

We can compare this expression with eq.~\eqref{eq:bcalD4d1_begin}. We follow the same procedure as in sec.~\ref{sec:OneCompactified}, the first component is -- both for large $L$ and small $L$ -- obtained after imposing the BPHZ subtraction and gives, 

\begin{equation}
\frac{1}{8\pi^2} - \frac{1}{16\pi^2} \ln \left(\frac{\ell^2}{M^2} + \sum_{j=1}^d\frac{(2\pi n_j^\ell + \pi \theta_j)^2}{M^2L_j^2}\right).
\end{equation}

\noindent With respect to the remaining, the dominant contribution from the sum over K-Bessel functions comes from the component $n_i=1$,
\begin{widetext}
\begin{multline}
\frac{4}{(4\pi)^2} \sqrt{\frac{\pi}{2}}
 \int_0^1dz \sum_{i=1}^d \cos\left(\pi \theta_i(1+z)\right)
\frac{e^{-L_i \sqrt{M^2+\left(\ell^2+\sum_{j=1}^d \frac{(2\pi n_j^\ell+\pi \theta_j)^2}{L_j^2}\right)z(1-z)}}}
{\sqrt{L_i \sqrt{M^2+\left(\ell^2+\sum_{j=1}^d \frac{(2\pi n_j^\ell+\pi \theta_j)^2}{L_j^2}\right)z(1-z)}}}\\
+
\frac{8}{(4\pi)^2} \sqrt{\frac{\pi}{2}}
\int_0^1dz \sum_{i_1=1}^d\sum_{i_2>i_1}^d \cos\left(\pi \theta_{i_1}(1+z)\right)
\cos\left(\pi \theta_{i_2}(1+z)\right)
\frac{e^{-\sqrt{L_{i_1}^2+L_{i_2}^2} \sqrt{M^2+\left(\ell^2+\sum_{j=1}^d \frac{(2\pi n_j^\ell+\pi \theta_j)^2}{L_j^2}\right)z(1-z)}}}
{\sqrt{\sqrt{L_{i_1}^2+L_{i_2}^2} \sqrt{M^2+\left(\ell^2+\sum_{j=1}^d \frac{(2\pi n_j^\ell+\pi \theta_j)^2}{L_j^2}\right)z(1-z)}}}
+ \ldots,
\end{multline}
Comparing with eq.~\eqref{eq:bcalD4d1_Kbessel} we see that we have a similar structure. Following the same steps we find that the asymptotic behavior is
\begin{multline}
\frac{\sqrt{2\pi}}{4\pi^2} 
\frac{1}{\ell^2+\sum_{j=1}^d\left(\frac{2\pi n_j^\ell+\pi \theta_j}{L_j}\right)^2}
\Bigg\{
\sum_{i=1}^d
\frac{\cos\left(\pi \theta_i \right)
	+
	\cos\left(2\pi \theta_i \right)
}{L_i^2} \Gamma\left(\frac{3}{2},ML_i\right)
\\+ 2
\sum_{i_1=1}^d
\sum_{i_2>i_1}^d
\frac{\cos\left(\pi \theta_{i_1} \right)\cos\left(\pi \theta_{i_2} \right)
	+
	\cos\left(2\pi \theta_{i_1} \right)\cos\left(2\pi \theta_{i_2} \right)
}{L_{i_1}^2+L_{i_2}^2} \Gamma\left(\frac{3}{2},M\sqrt{L_{i_1}^2+L_{i_2}^2}\right)
+ \ldots
\Bigg\}.
\end{multline}
\end{widetext}

Therefore, we have for the bubble-link in the scenario with $d$ compactified dimensions
\begin{multline}
\mathcal{B}^{4}_{d}(\ell;\omega^\ell_{n_\alpha})
\sim
\frac{1}{8\pi^2} - \frac{1}{16\pi^2} \ln \left(\frac{\ell^2}{M^2} + \sum_{j=1}^d\frac{(2\pi n_j^\ell + \pi \theta_j)^2}{M^2L_j^2}\right)
\\
+
\frac{1}{(2\pi)^{\frac{3}{2}}} 
\frac{1}{\ell^2+\sum_{j=1}^d\left(\frac{2\pi n_j^\ell+\pi \theta_j}{L_j}\right)^2} 
f_d(L_i,\theta_i),
\end{multline}
\begin{multline}
f_d(L_i,\theta_i) =\sum_{i=1}^d
\frac{\cos\left(\pi \theta_i \right)
	+
	\cos\left(2\pi \theta_i \right)
}{L_i^2} \Gamma\left(\frac{3}{2},ML_i\right)
\\+ 2
\sum_{i_1=1}^d
\sum_{i_2>i_1}^d
\frac{\cos\left(\pi \theta_{i_1} \right)\cos\left(\pi \theta_{i_2} \right)
	+
	\cos\left(2\pi \theta_{i_1} \right)\cos\left(2\pi \theta_{i_2} \right)
}{L_{i_1}^2+L_{i_2}^2} \times \\ \Gamma\left(\frac{3}{2},M\sqrt{L_{i_1}^2+L_{i_2}^2}\right)
+ \ldots
+ \\
2^{d-1}
\frac{\prod_{i=1}^d \cos(\pi\theta_i)+\prod_{i=1}^d \cos(2\pi\theta_i)}{\sum{i=1}^d L_i^2}
\Gamma\left(\frac{3}{2},M\sqrt{\sum_{i=1}^d L_i^2}\right).
\end{multline}

We substitute this back into eq.~\eqref{eq:bubblechain_regularized_borel} and obtain the sum over all bubble chains in the Borel plane as
\begin{multline}
B_{-\frac{\widehat{\mathcal{T}}}{g}}
\sim \frac{2P^4 e^{ -\frac{gu}{16\pi^2}} }{(4\pi)^{2-\frac{d}{2}} \Gamma\left(2-\frac{d}{2}\right)\prod_i L_i}
\sum_{n_i^\ell \in \mathbb{Z}^d}
\int_0^\infty d\ell\\
\frac{\ell^{3-d}}{(\ell^2+\omega_\ell^2+M^2)^3}
\left(\frac{M^2+\ell^2+\sum_{j=1}^d\frac{(2\pi n_j^\ell + \pi \theta_j)^2}{L_j^2}}{M^2}\right)^{\frac{gu}{32\pi^2}}
\\e^{ -\frac{gu}{2^{5/2}\pi^{3/2}}
	\frac{f_d(L_i,\theta_i)}{M^2+\ell^2+\sum_{j=1}^d\frac{(2\pi n_j^\ell + \pi \theta_j)^2}{L_j^2}} 
}.
\end{multline}
\noindent It has the same structure of the simplified scenario with $d=1$. Expanding the remaining exponential in power series and organizing the expression,
\begin{multline}
B_{-\frac{\widehat{\mathcal{T}}}{g}}
\sim \frac{2P^4 e^{ -\frac{gu}{16\pi^2}} }{(4\pi)^{2-\frac{d}{2}} \Gamma\left(2-\frac{d}{2}\right) M^{\frac{gu}{16\pi^2}} \prod_i L_i}
\sum_{k\in \mathbb{N}}\frac{1}{k!}\times\\
\left( -\frac{gu}{2^{5/2}\pi^{3/2}}f_d(L_i,\theta_i)\right)^k
\sum_{n_i^\ell \in \mathbb{Z}^d}
\int_0^\infty d\ell\\
\frac{\ell^{3-d}}{\left(M^2+\ell^2+\sum_{j=1}^d\frac{(2\pi n_j^\ell + \pi \theta_j)^2}{L_j^2}\right)^{3-\frac{gu}{32\pi^2}+k}}.
\end{multline}
\noindent Then we can compute the integral over the internal momentum $\ell$
\begin{multline}
B_{-\frac{\widehat{\mathcal{T}}}{g}}
\sim 
\frac{2P^4 e^{ -\frac{gu}{16\pi^2}} }{(4\pi)^{2-\frac{d}{2}} \Gamma\left(2-\frac{d}{2}\right) M^{\frac{gu}{16\pi^2}} \prod_i L_i}
\sum_{k\in \mathbb{N}}\frac{1}{k!}\times\\
\left( -\frac{gu}{2^{5/2}\pi^{3/2}}f_d(L_i,\theta_i)\right)^k
\sum_{n_i^\ell \in \mathbb{Z}^d}\\
\frac{1}{\left(M^2+\sum_{j=1}^d\frac{(2\pi n_j^\ell + \pi \theta_j)^2}{L_j^2}\right)^{1+\frac{d}{2}-\frac{gu}{32\pi^2}+k}}\times\\
\frac{\Gamma\left(\frac{4-d}{2}\right)\Gamma\left(1+\frac{d}{2}-\frac{gu}{32\pi^2}+k\right)}{2\Gamma\left(3-\frac{gu}{32\pi^2}+k\right)}.
\end{multline}

The sum with respect to the $d$ modes $n_i^\ell$ produces an Elizalde zeta function. It is known that the poles of the Elizalde zeta are all given by the gamma function in the first term\footnote{This affirmation is valid for $c\neq0$, see Eq.~\eqref{eq:footnote}.}. Therefore, we can reduce our analysis to just this first component,
\begin{multline}
B_{-\frac{\widehat{\mathcal{T}}}{g}}
\sim 
\frac{P^4 e^{ -\frac{gu}{16\pi^2}} }{(4\pi)^{2}}
\sum_{k\in \mathbb{N}}\frac{1}{k!}
\left( -\frac{gu}{2^{5/2}\pi^{3/2}}f_d(L_i,\theta_i)\right)^k
\times\\\left[
M^{-2-2k} 
\frac{\Gamma\left(1-\frac{gu}{32\pi^2}+k\right)}{\Gamma\left(3-\frac{gu}{32\pi^2}+k\right)}
+ \ldots
\right],
\end{multline}
	\noindent and we obtain that the structure of singularities in the Borel plane remains the same from the scenario with just one compactified dimension. There is a infinite countable set of poles located at the positive integers $\kappa_0u = j \in \mathbb{N}^*$ with $\kappa_0 = g/(2(4\pi)^2)$. What changes in the present scenario is just the value of the residues. It is plain evident that the location of the poles are independent of the size and the number of compactified dimensions $d$, the size of them $L_i$ and the twisted boundary condition $\theta_i$ affects only the residue of the poles. There are two special points: one at $L=\infty$ where some residues vanishes and we have just the poles from the bulk theory ($\mathbb{R}^D$);  and the scenario with $\theta_i=1$ where $f_d(L_i,\theta_i)=0$.

\section{About the Static mode approximation}
\label{sec:StaticMode}

In this section, we comment about a different and naive perspective. Here we choose to stick with the concept that we can achieve a dimensional reduction if we consider just the static mode, a naive application of the Appelquist-Carazzone decoupling theorem. Restricting ourselves to the static mode (zero mode) the bubble diagram with $d$ compactifications turns out to be
\begin{multline}
\mathcal{B}^{D}_{d}(\ell;\omega^\ell_{n_\alpha}) =
\frac{1}{\prod_{\alpha=1}^d L_\alpha} 
\frac{\Gamma\left[2 - \frac{D-d}{2}\right]}{(4\pi)^{\frac{D-d}{2}}} 
\int_{0}^{1}dz \\
\Bigg[ 
\sum_{\alpha=1}^{d} \frac{4\pi^2}{L_\alpha^2}\left(\frac{\theta_i}{2}(1+z)\right)^2 
+ M^2
\\+ \left(\ell^2
+ \sum_{i=1}^d \frac{(\pi \theta_i)^2}{L_i^2} \right) z(1-z)
\Bigg]^{-2+\frac{D-d}{2}},
\end{multline} 
\noindent or, if we take periodic boundary conditions (where there truly is a static mode) and consider $D=4$, we get
\begin{equation}
\mathcal{B}^{4}_{d}(\ell;\omega^\ell_{n_\alpha}) =
\frac{\Gamma\left[ \frac{d}{2}\right]}{(4\pi)^{\frac{4-d}{2}}L^d} 
\int_{0}^{1}dz 
\Bigg[ 
M^2
+ \ell^2 z(1-z)
\Bigg]^{-\frac{d}{2}}.
\end{equation} 

Which can be solved exactly for each number of compactified dimensions. Let us make the change of variables  $\ell^2\left(z-\frac{1}{2}\right)=\left(M^2+\frac{\ell^2}{4}\right)x^2$ so that
\begin{multline}
\mathcal{B}^{4}_{d}(\ell;\omega^\ell_{n_\alpha}) =
\frac{\Gamma\left[ \frac{d}{2}\right]}{(4\pi)^{\frac{4-d}{2}}L^d \ell} 
\left(M^2+\frac{\ell^2}{4}\right)^{\frac{1-d}{2}}\times\\
\int_{-c}^{+c}dx\; (1-x^2)^{-\frac{d}{2}},
\end{multline}
\noindent where $c = \frac{\ell/2}{\sqrt{M^2+\frac{\ell^2}{4}}} = (1+\frac{4M^2}{\ell^2})^{-\frac{1}{2}} \sim 1 - \frac{2M^2}{\ell^2}$.
For $d\ge2$ we consider $x= \tanh y$ and for $d=1$ we consider $x = \sin y$. Therefore,
\begin{equation}
\int_{-c}^{+c}dx\; (1-x^2)^{-\frac{d}{2}}
=
\begin{cases}
2 \arcsin(c), &d=1;\\
2 \text{arctanh}(c), &d=2;\\
2 \sinh(\text{arctanh}(c), &d=3;\\
\sinh(\text{arctanh}(c)) \times  &d=4.\\
\cosh(\text{arctanh}(c)) &\\+ \text{arctanh}(c),&
\end{cases}
\end{equation}

The asymptotic behavior for large values of $\ell$ is esasily obtained for each of them.  For $d=2$ we have $\text{arctanh}(c) = \frac{1}{2}\ln \frac{1+c}{1-c} \sim \ln \frac{\ell}{M}$. For $d=3$ it becames $\sinh(\text{arctanh}(c)) = \frac{c}{\sqrt{1-c^2}} \sim \frac{\ell}{2M}$. For $d=4$ the expression produces $\frac{c}{1-c^2}+\frac{1}{2}\ln \frac{1+c}{1-c} \sim \ln \frac{\ell}{M}+\frac{\ell^2}{4M^2}$. While for $d=1$ we have $2\arcsin(c) \sim \pi + 2\frac{M}{\ell}$. All these different asymptotic expressions for the bubble-link, when substituted back into the sum over the bubble-chain diagrams, do not produce any singularity in the Borel plane. Meaning that the model is renormalon-free in the static mode approximation. However, the static mode approximation is meaningless for our model. As discussed, there is a mixing over the contributions from static and dynamic modes and the decoupling theorem loses validity in our scenario once we introduced the bubble-chain contribution. 

We could also exhibit here that a ``partial" static mode approximation produces the same behavior (cancellation of the renormalon poles). That is, we could keep all modes to compute the bubble-link contribution but consider just the static mode when computing the sum over the bubble-chain diagrams. With this, the argument of the K-Bessel function would go to zero in the limit of small $L$, and the logarithmic behavior cancels out algebraically. However, this would produce an unnecessarily large amount of computations and no new useful information at all. That is why we choose to show this simplified scenario of the full static mode approximation.

\section{Conclusions}
\label{sec:conclusions}

We investigated the behavior of the renormalon singularities in a toy model (scalar field theory) with the influece of spatial compactifications and boundary conditions. We considered the set of bubble-chain diagrams. The bubble-link amplitude produces a logarithmic behavior, and the sum over the bubble-chain is divergent. Looking at the Borel plane, we were able to investigate the source of the divergence of the asymptotic series: renormalon singularities at the positive real axis. Although we expected the renormalon poles to suffer some influence with respect to the many compactifications, we found that in the compactified model ($\mathbb{R}^D \rightarrow \mathbb{R}^{D-d}\times \mathbb{S}^d$) the renormalon singularities are just a set of poles located at the positive integer values of $\kappa_0 u$ (with $\kappa_0 = g/(2(4\pi)^2))$). The position and quantity of the poles are independent of the number and size of the compactified dimensions. The only modification is that there is a finite set of poles at $L=\infty$ (the original $\mathbb{R}^D$ scenario), and for finite $L$, we have an infinite (countable) set of poles.

We remark that we can obtain some cancellation of renormalons in the limits of the bulk, in concordance with the $L=\infty$ scenario, and also when we consider antiperiodic boundary conditions in space ($\theta=1$). Both scenarios make the residues of some poles vanish. Concerning the residues, they keep information about the length scale, the number of compactified dimensions, and the boundary condition. 

Regarding the structure of poles, our example serves to illustrate that one must be very careful with the asymptotic approximation employed. Say, for example, that we take $\ln \ell^2+M^2$ as $\ln \ell^2$ for large $\ell$, it can modify the argument of the gamma functions that appear due to the integral over $\ell$. This inocent change alters the location of the poles. Therefore, we must stick to an approximation that keeps more information about the original function. That is why we keep the $M^2$, both as a natural IR cutoff and a reminder. Based on this simple discussion, we enforce the need to carefully check any result in the literature introducing new renormalon poles as it might be a consequence not of the model but from the asymptotic approximation employed.

We must also remark that the observation that the renormalon poles are independent of the size is \textit{not} a general property independent of the model. For example, ref.~\cite{Ashie:2020bvw} presents a scenario where a small size modifies the structure. In the language of the present article, we can understand it by inspecting the argument of the K-Bessel function. In our scenario, we exhibit that the argument of the K-Bessel function is always large, regardless of the length scale $L$. In ref.~\cite{Ashie:2020bvw}, on the other hand, the argument of the K-Bessel goes to zero in the regime of small sizes (near dimensional reduction), and it requires a different approximation. It is possible to show, both following the procedure of ref.~\cite{Ashie:2020bvw} or by reexpressing the Elizalde zeta function in this new regime, that the logarithmic behavior vanishes completely. It remains, however, the inquiry of which models have renormalons singularities independent of size effects and which models depend on the number or size of the compactifications. 

\acknowledgments{The author thanks the Brazilian agency Conselho Nacional de Desenvolvimento Cient\'ifico e Tecnol\'ogico (CNPq) for financial support.}

\appendix

\bibliography{Criteria_Bubble5_Allways}{}
\bibliographystyle{apsrev4-1}


\end{document}